\documentclass{amsart}
\usepackage{amsfonts}
\usepackage{amsmath}
\usepackage{amssymb}
\usepackage{graphicx}

\begin{document}
\title[Reference Potential Approach to the Inverse Problem: II]{Reference potential approach to the quantum-mechanical inverse problem:
\linebreak II. Solution of Krein equation}
\author{Matti Selg}
\address{Riia 142, 51014 Tartu, Estonia}
\email{matti@fi.tartu.ee}

\begin{abstract}
A reference potential approach to the one-dimensional quantum-mechanical
inverse problem is developed. All spectral characteristics of the system,
including its discrete energy spectrum, the full energy dependence of the
phase shift, and the Jost function, are expected to be known. The technically
most complicated task in ascertaining the potential, solution of a relevant
integral equation, has been decomposed into two relatively independent
problems. First, one uses Krein method to calculate an auxiliary potential
with exactly the same spectral density as the initial reference potential, but
with no bound states. Thereafter, using Gelfand-Levitan method, it is possible
to introduce, one by one, all bound states, along with calculating another
auxiliary potential of the same spectral density at each step. For the system
under study (diatomic xenon molecule), the kernel of the Krein integral
equation can be accurately ascertained with the help of solely analytic means.
At small distances the calculated auxiliary potential with no bound states
practically coincides with the initial reference potential, which is in full
agreement with general theoretical considerations. Several possibilities of
solving the Krein equation are proposed and the prospects of further research discussed.

\end{abstract}
\maketitle

\section{Introduction}

In this paper, a recently proposed reference potential approach to the
one-dimensional quantum mechanical inverse problem \cite{SelgI} is further
developed. Let us briefly recall that the starting idea is to choose a
suitable reference potential for the system. For this fixed potential, it is
always possible to calculate all its spectral characteristics, which means
that in this artificial way one can obtain the complete set of information
(otherwise inaccessible) needed to uniquely solve the inverse problem: 1) full
energy spectrum of the bound states; 2) full energy dependence of the phase
shift for the scattering states; 3) norming constants of the regular energy
eigenfunctions for all bound states. Of course, there is no need to regain the
potential which is already known by definition. However, the quantities
related to the reference potential can be used as zeroth approximations to the
real spectral characteristics of the system. A further step might be, for
example, to calculate another potential whose discrete energy levels would
exactly fit with their actually observed values, so that at least in this
sense the new potential would be more realistic than the initial reference potential.

To illustrate the method, as previously, Xe$_{2}$ molecule is under study, and
the same three-component exactly solvable reference potential is used. We have
already ascertained the full energy dependence of the phase shift for the
scattering states \cite{SelgI} and demonstrated its excellent agreement with
the celebrated Levinson theorem \cite{Levinson}. In addition, we are provided
with full knowledge of the Jost function, which is the most important spectral
characteristic of the system. Thus, we are prepared to attack the most serious
computational-technical problem, solution of an integral equation, which would
enable to uniquely ascertain the potential. For this purpose a combined
approach is used. In Section 2 we recall some useful properties of the
Gelfand-Levitan method \cite{GL}, which enable to separate the main problem
into two independent parts: 1) calculating an auxiliary potential $V_{0}(r)$
with exactly the same spectral density for positive energies as the reference
potential $V(r)$, but with no bound states; 2) calculating a series of
auxiliary potentials $V_{k}(r),$ $k=1,2,...,n$ (i.e., adding, one by one, all
bound states, and keeping their norming constants), also having the same
spectral density for positive energies. The second step is, in fact, less
complicated and is only briefly discussed in this paper. Our main goal is to
ascertain the auxiliary potential $V_{0}(r)$ with no bound states, starting
from the known Jost function for positive energies. This is the subject of
Section 3 where the Krein method \cite{Krein1, Krein2} will be used. Some
concluding remarks and a discussion of the further perspectives of the method
form the content of Section 4.

\section{Step-by-step use of Gelfand-Levitan method}

First, let us recall general solution scheme in the frame of Gelfand-Levitan
method \cite{GL} (see, e.g., \cite{Chadan} for more details). One's aim is to
solve the integral equation%
\begin{equation}
K(r,r^{\prime})+G(r,r^{\prime})+\int\limits_{0}^{r}K(r,s)G(s,r^{\prime})ds=0,
\end{equation}
whose kernel%
\begin{equation}
G(r,r^{\prime})=\int\limits_{-\infty}^{\infty}\frac{\sin\left(  kr\right)
\cdot\sin\left(  kr^{\prime}\right)  }{k^{2}}d\sigma,
\end{equation}
is determined by the quantity $d\sigma\equiv\left[  \dfrac{d\rho_{2}(E)}%
{dE}-\dfrac{d\rho_{1}(E)}{dE}\right]  dE,$ which contains the the difference
of two spectral densities%
\begin{equation}
\dfrac{d\rho(E)}{dE}=\left\{\genfrac{}{}{0pt}{}{\pi^{-1}\sqrt{E}\left\vert 
F(E)\right\vert ^{-2},\text{{}}E\geq0,}{\sum\limits_{n}C_{n}\delta(E-E_{n}),
\text{ \ }E<0,}\right.
\end{equation}
one of them ($\dfrac{d\rho_{1}(E)}{dE}$) being related to a known potential
($V_{1}(r)$). In this sense Gelfand-Levitan method also represents a reference
potential approach. The simplest possibility is to take $V_{1}(r)\equiv0$, and
correspondingly, $\dfrac{d\rho_{1}(E)}{dE}=\dfrac{\sqrt{E}}{\pi}$. Then%
\begin{equation}
G(r,r^{\prime})=\frac{2}{\pi}\int\limits_{0}^{\infty}\sin\left(  kr\right)
\cdot\sin\left(  kr^{\prime}\right)  g(k)dk+\sum_{n}\frac{C_{n}}{4\gamma
_{n}^{2}}\sinh\left(  \gamma_{n}r\right)  \sinh\left(  \gamma_{n}r^{\prime
}\right)  ,
\end{equation}
where $\gamma_{n}^{2}=-\dfrac{2mE_{n}}{\hbar^{2}}$, $E_{n}$ being the bound
levels and $C_{n}$, their related norming constants. The characteristic
function%
\begin{equation}
g(k)\equiv\dfrac{1}{\left\vert F(k)\right\vert ^{2}}-1
\end{equation}
is determined by the modulus of the Jost function $F(k)$. Thus, in principle,
one can solve Eq. (1) and then calculate the potential%
\begin{equation}
V(r)=2C\frac{d}{dr}K(r,r),\text{ }C\equiv\frac{\hbar^{2}}{2m}.\text{ }%
\end{equation}

This scheme might seem simple, but its actual realization stumbles upon
serious computational-technical difficulties. Now, let us assume that we
somehow managed to ascertain an auxiliary potential $V_{0}(r)$ with exactly
the same spectral density for positive energies as the desired potential
$V(r)$, but having no bound states. In this case one easily finds that
\cite{Chadan}%
\begin{equation}
G(r,r^{\prime})=\sum_{j}C_{j}\varphi_{0}\left(  i\gamma_{j},r\right)
\varphi_{0}\left(  i\gamma_{j},r^{\prime}\right)  ,
\end{equation}
where $\varphi_{0}\left(  i\gamma_{j},r\right)  $ are the regular solutions
(NB! not the real eigenfunctions!) related to the auxiliary potential
$V_{0}(r).$ Next, let us introduce another auxiliary potential $V_{1}(r)$ with
just one bound eigenvalue $E_{1}$ (note that numeration of the levels here
starts from the highest-energy one, and therefore, $E_{n}$ corresponds to the
zeroth level), and again, with exactly the same spectral density for positive
energies ($\left\vert F(k)\right\vert $ remains the same). Thereafter, one
introduces an auxiliary potential $V_{2}(r)$ with two levels ($E_{1}$ and
$E_{2}$), etc., until he comes to the desired potential $V_{n}(r)=V(r)$.

It can be proved that%
\begin{equation}
V_{j}(r)-V_{0}(r)=-2C\left\{  \ln\left[  \det C_{j}(r)\right]  \right\}
^{\prime\prime},\text{ }j=1,2,...,n
\end{equation}
and the corresponding regular solution
\begin{equation}
\varphi_{j}\left(  k,r\right)  =\det\left\vert
\genfrac{}{}{0pt}{}{C_{j}(r)\text{ \ \ }\Psi_{j}(r)}{\beta_{j}(k,r)\text{
\ }\varphi_{0}\left(  k,r\right)  }\right\vert \left[  \det C_{j}(r)\right]  
^{-1}.
\end{equation}
Here, a $j\times j$ matrix $C_{j}(r)\equiv I+\int\limits_{0}^{r}R^{(j)}(s)ds$
($I$ is the unit matrix) with the elements of $R^{(j)}$ being
\begin{equation}
R_{lm}^{(j)}=C_{l}\varphi_{0}\left(  i\gamma_{l},r\right)  \varphi_{0}\left(
i\gamma_{m},r\right)  ,\text{ (}l,m=1,2,...,j\text{)}%
\end{equation}
a column vector (do not confuse the norming constants $C_{l}$ with the
matrices $C_{j}(r)$)
\begin{equation}
\Psi_{j}(r)=\left(\genfrac{}{}{0pt}{}{\genfrac{}{}{0pt}{0}{C_{1}\varphi_{0}
\left(  i\gamma_{1},r\right)  }{C_{2}\varphi_{0}\left(  i\gamma_{2},r\right)  }%
}{\genfrac{}{}{0pt}{0}{\cdot}{\genfrac{}{}{0pt}{0}{\cdot
}{\genfrac{}{}{0pt}{0}{\cdot}{C_{j}\varphi_{0}\left(  i\gamma_{j},r\right)  }%
}}}\right)  ,
\end{equation}
and a row vector%
\begin{equation}
\beta_{j}(k,r)=\left(  \int\limits_{0}^{r}\varphi_{0}\left(  i\gamma
_{1},s\right)  \varphi_{0}\left(  k,s\right)  ds\text{ \ }\int\limits_{0}%
^{r}\varphi_{0}\left(  i\gamma_{2},s\right)  \varphi_{0}\left(  k,s\right)
ds\text{ ...}\int\limits_{0}^{r}\varphi_{0}\left(  i\gamma_{j},s\right)
\varphi_{0}\left(  k,s\right)  ds\right)
\end{equation}
have been introduced.

To be more specific, let us examine the simplest case of just one bound state
($j=1$). Then $C_{j}(r)$ reduces to scalar and one gets%
\begin{equation}
V_{1}(r)=V_{0}(r)-2C\left\{  \ln\left[  1+C_{1}\int\limits_{0}^{r}\varphi
_{0}^{2}\left(  i\gamma_{1},s\right)  ds\right]  \right\}  ^{\prime\prime}.
\end{equation}
As can be shown \cite{Chadan}, the regular solution $\varphi_{1}\left(
i\gamma_{1},r\right)  $ related to $V_{1}(r)$ (i.e., the real confined
eigenfunction) reads%

\begin{equation}
\varphi_{1}\left(  i\gamma_{1},r\right)  =\frac{\varphi_{0}\left(  i\gamma
_{1},r\right)  }{1+C_{1}\int\limits_{0}^{r}\varphi_{0}^{2}\left(  i\gamma
_{1},s\right)  ds}.
\end{equation}
Let us prove that $C_{1}$, as needed, is the norming constant of $\varphi
_{1}\left(  i\gamma_{1},r\right)  $. Indeed, slightly rearranging Eq. (14):
$1+C_{1}\int\limits_{0}^{r}\varphi_{0}^{2}\left(  i\gamma_{1},s\right)
ds=\dfrac{\varphi_{0}\left(  i\gamma_{1},r\right)  }{\varphi_{1}\left(
i\gamma_{1},r\right)  }$, differentiating both sides: $C_{1}\varphi_{0}%
^{2}\left(  i\gamma_{1},r\right)  =\dfrac{\varphi_{0}^{\prime}\varphi
_{1}-\varphi_{0}\varphi_{1}^{\prime}}{\varphi_{1}^{2}},$ i.e., $C_{1}%
\varphi_{1}^{2}\left(  i\gamma_{1},r\right)  =-\left(  \frac{1}{\varphi_{0}%
}\right)  ^{\prime}\varphi_{1}-\dfrac{1}{\varphi_{0}}\varphi_{1}^{\prime
}=-\left(  \dfrac{\varphi_{1}}{\varphi_{0}}\right)  ^{\prime}$, and
integrating both sides of the latter equation, one gets $\int\limits_{0}%
^{\infty}C_{1}\varphi_{1}^{2}\left(  i\gamma_{1},r\right)  dr=1$, because any
regular solution $\varphi(r)\approx r$ as $r\rightarrow0$, and therefore,
$\lim_{r\rightarrow0}\left(  \dfrac{\varphi_{1}}{\varphi_{0}}\right)  =1$.

Now, let us analyze a more complicated case of two bound states ($j=1$).
According to Eqs. (8) and (13),
\begin{equation}
V_{2}(r)-V_{1}(r)=V_{2}(r)-V_{0}(r)-\left[  V_{1}(r)-V_{0}(r)\right]
=-2C\left\{  \ln\left[  \frac{\det C_{2}(r)}{1+C_{1}\int\limits_{0}^{r}%
\varphi_{0}^{2}\left(  i\gamma_{1},s\right)  ds}\right]  \right\}
^{\prime\prime}.
\end{equation}
Here, the $2\times2$ matrix $C_{2}(r)=\left(\genfrac{}{}{0pt}{0}{C_{11}(r)
\text{ \ }C_{12}(r)}{C_{21}(r)\text{ \ }C_{22}(r)}\right)  $ has the elements
\begin{gather}
C_{11}(r)=1+C_{1}\int\limits_{0}^{r}\varphi_{0}^{2}\left(  i\gamma
_{1},s\right)  ds,\text{ }C_{12}(r)=C_{1}\int\limits_{0}^{r}\varphi_{0}\left(
i\gamma_{1},s\right)  \varphi_{0}\left(  i\gamma_{2},s\right)  ds\\
C_{21}(r)=\frac{C_{2}}{C_{1}}C_{12}(r),\text{ }C_{22}(r)=1+C_{2}%
\int\limits_{0}^{r}\varphi_{0}^{2}\left(  i\gamma_{2},s\right)  ds,\nonumber
\end{gather}
and therefore, the argument of the logarithm in Eq. (15) reads%
\begin{gather}
f(r)\equiv C_{22}-\frac{C_{1}C_{2}\left[  \int\limits_{0}^{r}\varphi
_{0}\left(  i\gamma_{1},s\right)  \varphi_{0}\left(  i\gamma_{2},s\right)
ds\right]  ^{2}}{1+C_{1}\int\limits_{0}^{r}\varphi_{0}^{2}\left(  i\gamma
_{1},s\right)  ds}=\\
=1+C_{2}\left\{  \int\limits_{0}^{r}\varphi_{0}^{2}\left(  i\gamma
_{2},s\right)  ds-\frac{C_{1}\left[  \beta_{12}(r)\right]  ^{2}}{1+C_{1}%
\int\limits_{0}^{r}\varphi_{0}^{2}\left(  i\gamma_{1},s\right)  ds}\right\}
,\nonumber
\end{gather}
where $\beta_{12}(r)\equiv\int\limits_{0}^{r}\varphi_{0}\left(  i\gamma
_{1},s\right)  \varphi_{0}\left(  i\gamma_{2},s\right)  ds$.

As can be easily proved,%
\begin{equation}
f(r)=1+C_{2}\int\limits_{0}^{r}\varphi_{1}^{2}\left(  i\gamma_{2},s\right)
ds.
\end{equation}
Indeed, using Eq. (14) and a more general formula \cite{Chadan}%
\[
\varphi_{1}\left(  k,r\right)  =\varphi_{0}\left(  k,r\right)  -C_{1}%
\varphi_{0}\left(  i\gamma_{1},r\right)  \times\frac{\int\limits_{0}%
^{r}\varphi_{0}\left(  i\gamma_{1},s\right)  \varphi_{0}\left(  k,s\right)
ds}{1+C_{1}\int\limits_{0}^{r}\varphi_{0}^{2}\left(  i\gamma_{1},s\right)
ds},
\]
and taking $k=i\gamma_{2}$, one gets%
\[
\varphi_{1}\left(  i\gamma_{2},r\right)  =\varphi_{0}\left(  i\gamma
_{2},r\right)  -C_{1}\varphi_{1}\left(  i\gamma_{1},r\right)  \beta_{12}(r),
\]
i.e.,%
\begin{align*}
\varphi_{1}^{2}\left(  i\gamma_{2},r\right)   &  =\varphi_{0}^{2}\left(
i\gamma_{2},r\right)  +\left[  C_{1}\varphi_{1}\left(  i\gamma_{1},r\right)
\beta_{12}(r)\right]  ^{2}-2C_{1}\varphi_{0}\left(  i\gamma_{2},r\right)
\varphi_{1}\left(  i\gamma_{1},r\right)  \beta_{12}(r)=\\
&  =\varphi_{0}^{2}\left(  i\gamma_{2},r\right)  -C_{1}\left[  \frac
{\varphi_{1}\left(  i\gamma_{1},r\right)  }{\varphi_{0}\left(  i\gamma
_{1},r\right)  }\beta_{12}^{2}(r)\right]  ^{\prime},
\end{align*}
which, after integrating from 0 to $r$, and taking account of Eq. (14), proves
the claim.

Thus, in full analogy with Eq. (13), one can write%
\begin{equation}
V_{2}(r)=V_{1}(r)-2C\left\{  \ln\left[  1+C_{2}\int\limits_{0}^{r}\varphi
_{1}^{2}\left(  i\gamma_{2},s\right)  ds\right]  \right\}  ^{\prime\prime},
\end{equation}
and more generally,
\begin{equation}
V_{k}(r)=V_{k-1}(r)-2C\left\{  \ln\left[  1+C_{k}\int\limits_{0}^{r}%
\varphi_{k-1}^{2}\left(  i\gamma_{k},s\right)  ds\right]  \right\}
^{\prime\prime},\text{ }k=1,2,...,n.
\end{equation}
This way one gets rid of inconvenient matrix equations and can move towards
the goal step-by-step, introducing one bound state at a time.

The described procedure can be reverted \cite{AM, LP}, i.e., starting from the
potential $V_{k}(r)$, one can construct the auxiliary potential $V_{k-1}(r)$,
removing the zeroth level ($E_{k}$ according to the numeration used). The new
potential reads%
\begin{equation}
V_{k-1}(r)=V_{k}(r)+2C\left\{  \frac{2\Psi_{0}^{(k)}(r)\left(  \Psi_{0}%
^{(k)}(r)\right)  ^{\prime}}{\int\limits_{r}^{\infty}\left(  \Psi_{0}%
^{(k)}(s)\right)  ^{2}ds}+\left[  \frac{\left(  \Psi_{0}^{(k)}(r)\right)
^{2}}{\int\limits_{r}^{\infty}\left(  \Psi_{0}^{(k)}(s)\right)  ^{2}%
ds}\right]  ^{2}\right\}  ,\text{ (}k=n,n-1,...,1\text{)}%
\end{equation}
where $\Psi_{0}^{(k)}(r)$ is the eigenfunction of the zeroth level, not
necessarily normalized (note that the norming constant is absent here). One
can remove, one by one, all bound states until he comes to the potential
$V_{0}(r)$ with no bound states. An illustration to Eq. (21) can be seen in
Fig. 1.

Thus, we have demonstrated that there is no need to solve Eq. (1) all at once.
It is probably much easier to first solve this equation for the auxiliary
potential $V_{0}(r)$ and then, step-by-step, introduce the bound states as
described above. The Jost functions $F_{n}(k)$ and $F_{0}(k)$ related to the
potentials $V_{n}(r)$ and $V_{0}(r)$, respectively, are connected by a simple
formula%
\begin{equation}
F_{n}(k)=F_{0}(k)\prod\limits_{j=1}^{n}\left(  \frac{k-i\gamma_{j}}%
{k+i\gamma_{j}}\right)  ,
\end{equation}
which demonstrates that these potentials, indeed, have exactly the same
spectral densities for positive energies . On the other hand, Eq. (22)
illustrates the general rule that any zero of the Jost function corresponds to
a bound state.

To end this section, let us recall important asymptotic formulas (see
\cite{Chadan} for details)%
\begin{equation}
V_{n}(r)-V_{0}(r)\approx-4C\left(  \sum_{j=1}^{n}C_{j}\right)  r,\text{
}r\rightarrow0,
\end{equation}%
\begin{equation}
V_{n}(r)-V_{0}(r)\approx-\frac{2C}{C_{1}}\left(  2\gamma_{1}\right)  ^{5}%
\exp(-2\gamma_{1}r),\text{ }r\rightarrow\infty.
\end{equation}
From Eq. (23) one can infer that in the immediate vicinity of the zero point
$r=0$ the potentials $V_{n}(r)$ and $V_{0}(r)$ practically coincide. Indeed,
all norming constants $C_{j}=\left[  \int\limits_{0}^{\infty}\varphi_{j}%
^{2}\left(  i\gamma_{j},r\right)  dr\right]  ^{-1}$are extremely small
quantities, since they are related to the regular solutions proportional to
$r$ as $r\rightarrow0$, and vanishing as $r\rightarrow\infty$, but achieving
very large absolute values between these asymptotic regions.

\section{Krein method}

Thus, in view of the results of previous section, our main goal is to
accurately ascertain the auxiliary potential $V_{0}(r)$ with no bound states,
provided that its spectral density $\dfrac{d\rho_{0}(E)}{dE}=\pi^{-1}\sqrt
{E}\left\vert F(E)\right\vert ^{-2}$ is known. This is indeed the case, since
we have carefully calculated the modulus of the Jost function for the
reference potential $V(r)$ \cite{SelgI}, and the whole idea is to use the same
quantity ($\left\vert F(E)\right\vert $) for the auxiliary potential
$V_{0}(r)$ as well. Since there are no bound states, the kernel of the
Gelfand-Levitan equation, according to Eq. (4), becomes analytically very
simple:%
\begin{equation}
G(r,r^{\prime})=H(r-r^{\prime})-H(r+r^{\prime}),
\end{equation}
where a new function%
\begin{equation}
H(r)\equiv\pi^{-1}\int\limits_{0}^{\infty}g(k)\cos(kr)dk,
\end{equation}
with $g(k)$ given by Eq. (5), has been introduced. The function $H(r)=H(-r)$
is very important for the further treatment, wherefore a special designation,
Krein $H$-function, will be used for this item henceforward. A motivation for
this name stems from a very useful method for solving the inverse problem,
which has been elaborated by Krein \cite{Krein1, Krein2}. This method is based
on a Fredholm-type integral equation%
\begin{equation}
\Gamma_{2r}(r^{\prime})+H(r^{\prime})+\int\limits_{0}^{2r}\Gamma
_{2r}(s)H(s-r^{\prime})ds=0.
\end{equation}
It can be shown that the solutions of Gelfand-Levitan and Krein equations are
linked by a simple formula (cf. with Eq. (25))
\begin{equation}
K(r,r^{\prime})=\Gamma_{2r}(r-r^{\prime})-\Gamma_{2r}(r+r^{\prime}),
\end{equation}
and the desired potential reads%
\begin{equation}
V_{0}(r)=4C\left\{  \left[  G(x)\right]  ^{2}-\frac{dG(x)}{dx}\right\}
,\text{ }x\equiv2r,
\end{equation}
where $G(x)\equiv\Gamma_{2r}(2r).$ A special notation, Krein $G$-function,
will used for this quantity henceforward.

Without any doubt, Krein method is well suited for our purposes, but there is
still a lot of analytical and computational-technical work to do. First, we
have to accurately ascertain the Krein $H$-function, which, according to Eq.
(26), is simply the Fourier cosine transform of the characteristic function
$g(k)$ whose full energy dependence has already been ascertained \cite{SelgI}.
One might think that the $H$-function can be determined using the well-known
fast Fourier transform technique. This, however, is an erroneous view, because
(as we demonstrate below) the $H$-function has to be calculated in a wide
distance range with very small step, to ensure the correct asymptotic behavior
of the resulting potential. Fortunately, the problem can be solved accurately
and quite easily with the help of solely analytic means.

\subsection{Calculation of $H$-function}

The function $g(k)$ is shown in Fig. 2. In addition to the overall curve, some
characteristic slices can be seen in the insets. There exists a range
$k\in(0,k_{1})$ where $g(k)=-1$ (with high accuracy), and for this range one
immediately gets the relevant component of the $H$-function%
\[
H_{0}(r)=-\frac{\sin(k_{1}r)}{\pi r}.
\]
Now, let us see how the $g(k)$ curve passes through the "critical" region
around $k_{0}=\sqrt{V(0)/C}$. As it happens (see the upper inset in Fig. 2),
in a narrow range $k\in(k_{1},k_{2})$ ($k_{0}$ is also located within this
range) $g(k)+1$ can be nicely approximated by a sum of several Gaussians:%

\begin{equation}
g(k)=-1+\sum_{j}a_{j}\exp\left[  -\frac{1}{2}\left(  \frac{k-\widetilde{k}%
_{j}}{b_{j}}\right)  ^{2}\right]  ,
\end{equation}
where the parameters $a_{j},b_{j}$ and $\widetilde{k}_{j}$ can be determined
from a least-squares fit. For the reference potential examined here, four such
components have been introduced, and the value $k_{2}=$ 19230 \AA $^{-1}$ has
been chosen. Thus, in the above formula one can replace $k_{1}$ with $k_{2}$
(due to -1 in Eq. (30)), i.e.,%
\begin{equation}
H_{1}(r)=-\frac{\sin(k_{2}r)}{\pi r},
\end{equation}
while the sum of Gaussians gives another component of the Krein $H$-function,
$H_{2}(r)=$ $\sum\limits_{j}H_{2}^{(j)}(r),$ where all constituents
$H_{2}^{(j)}(r)$ can be ascertained analytically. Indeed, one can introduce a
new independent variable $x\equiv\dfrac{k-\widetilde{k}_{j}}{\sqrt{2}b_{j}},$
and calculate%
\begin{gather}
H_{2}^{(j)}(r)=\frac{\sqrt{2}a_{j}b_{j}}{\pi}\int\limits_{x_{1}}^{x_{2}}%
\exp(-x^{2})\cos\left(  \sqrt{2}b_{j}xr+\widetilde{k}_{j}r\right)
dx=\frac{a_{j}b_{j}}{\sqrt{2\pi}}\exp(-\frac{1}{2}b_{j}^{2}r^{2})\times\\
\times\left\{  \cos(\widetilde{k}_{j}r)\left[  \operatorname{Re}%
\operatorname{erf}(y_{2})-\operatorname{Re}\operatorname{erf}(y_{1})\right]
-\sin(\widetilde{k}_{j}r)\left[  \operatorname{Im}\operatorname{erf}%
(y_{2})-\operatorname{Im}\operatorname{erf}(y_{1})\right]  \right\}
,\nonumber
\end{gather}
where $x_{l}\equiv\dfrac{k_{l}-\widetilde{k}_{j}}{\sqrt{2}b_{j}}$ and
$y_{l}=x_{l}-\dfrac{ib_{j}r}{\sqrt{2}}$($l=1,2$ and $i$ is the imaginary
unit). The error function, for any complex argument $z,$ can be evaluated in
terms of the confluent hypergeometric functions \cite{Bateman}%
\begin{equation}
\operatorname{erf}(z)=\frac{\sqrt{2}}{\pi}z\Phi\left(  \frac{1}{2},\frac{3}%
{2};-z^{2}\right)  ,
\end{equation}
where $\Phi(a,c;x)=1+\dfrac{ax}{1!c}+\dfrac{a(a+1)x^{2}}{2!c(c+1)}+...$ For
large arguments another expression is more convenient:%
\begin{equation}
\operatorname{erf}(z)=1-\frac{\exp(-z^{2})}{\sqrt{\pi}z}\left[  1-\frac
{1}{2z^{2}}+\frac{1\cdot3}{\left(  2z^{2}\right)  ^{2}}-\frac{1\cdot3\cdot
5}{\left(  2z^{2}\right)  ^{3}}+...\right]  .
\end{equation}
Consequently, any constituent of the Krein $H$-function expressed by Eq. (32)
can be easily ascertained with any desired accuracy.

Next one can introduce an arbitrary (but still reasonable) boundary point
$k_{3}$, and approximate $g(k)$ in the range $k\in(k_{2},k_{3})$ as follows:%
\begin{equation}
g(k)=\sum_{j}a_{j}\exp\left[  -b_{j}(k-\widetilde{k}_{j})\right]  .
\end{equation}
This brings along another component of the $H$-function $H_{3}(r)=$
$\sum\limits_{j}H_{3}^{(j)}(r)$ with constituents%
\begin{gather}
H_{3}^{(j)}(r)=\frac{a_{j}}{\pi}\int\limits_{k_{2}}^{k_{3}}\exp\left[
-b_{j}(k-\widetilde{k}_{j})\right]  \cos\left(  kr\right)  dk=\frac{a_{j}%
\exp(b_{j}\widetilde{k}_{j})}{\pi\left(  b_{j}^{2}+r^{2}\right)  }\times\\
\times\left\{  \exp(-b_{j}k_{3})\left[  r\sin\left(  k_{3}r\right)  -b_{j}%
\cos\left(  k_{3}r\right)  \right]  -\exp(-b_{j}k_{2})\left[  r\sin\left(
k_{2}r\right)  -b_{j}\cos\left(  k_{2}r\right)  \right]  \right\}  .\nonumber
\end{gather}
Introducing new suitable boundary points $k_{4}$, $k_{5}$, etc., the
approximation of $g(k)$ in the form of Eq. (35) can be continued until the
conventional starting point $k_{a}$ of the asymptotic region (see the
treatment below). The number of these boundary points, as well as the number
of exponents in any particular interval, is, of course, a subject for probes
and trials. We have introduced four such intervals and used a different
three-exponent approximation in any of them. As has been carefully checked,
this ensures the accuracy of at least 6 significant digits for the calculated
Krein $H$-function in the whole physical domain. The parameters of all
components are given in Table 1.

For the remaining part of the $k$-space the asymptotic formula%
\begin{equation}
\ln\left\vert F(k)\right\vert =\dfrac{a_{2}}{k^{2}}+\frac{a_{4}}{k^{4}}%
+\frac{a_{6}}{k^{6}}+...,\text{ }k\geq k_{a},
\end{equation}
can be used, where \cite{SelgI}%
\begin{equation}
a_{2}=\frac{V(0)}{4C},\text{ }a_{4}=\frac{2\left[  V(0)\right]  ^{2}%
-CV^{\prime\prime}(0)}{16C^{2}}.
\end{equation}
The coefficient $a_{6},$ as well as the coefficients for higher-order terms,
can also be calculated in terms of the reference potential and its
derivatives, but the resulting expressions are rather complicated and
inconvenient for practical use. Instead, we only introduced just one
additional term, $\dfrac{a_{6}}{k^{6}}$, and determined the coefficient
$a_{6}$ from the general demand (see Eq. (23)) that the potentials $V_{0}(r)$
and $V(r)$ should coincide as $r\rightarrow0$. Such physically well motivated
choice of $a_{6}$ is indeed possible, as will be explained below. Thus, within
this approximation, the asymptotic part of $g(k)$ reads%
\begin{equation}
g(k)=\dfrac{b_{1}}{k^{2}}+\frac{b_{2}}{k^{4}}+\frac{b_{3}}{k^{6}},\text{
}k\geq k_{a},
\end{equation}
where%
\begin{equation}
b_{1}=-2a_{2}\text{, }b_{2}=-2(a_{4}-a_{2}^{2})\text{, }b_{3}=-2(a_{6}%
-2a_{2}a_{4}+\frac{2}{3}a_{2}^{3}).
\end{equation}
The relevant asymptotic component of the $H$-function becomes%
\[
H_{a}(r)\equiv\pi^{-1}\int\limits_{k_{a}}^{\infty}g(k)\cos(kr)dk.
\]

Since the expression for $H_{a}(r)$ will contain the sine integral
$\operatorname{Si}(x_{a})\equiv\int\limits_{0}^{x_{a}}\dfrac{\sin t}{t}dt,$
let us recall a useful formula \cite{Bateman}%
\begin{equation}
\operatorname{Si}(x)=\frac{\pi}{2}-\frac{i}{2}\exp(-ix)\Psi(1,1;ix)+\frac
{i}{2}\exp(ix)\Psi(1,1;-ix),
\end{equation}
where $i$ is the imaginary unit, and the function $\Psi(a,c;z)$ is a
particular solution of the confluent hypergeometric equation introduced by
Tricomi. For a large argument it can be evaluated from the asymptotic series
\begin{equation}
\Psi(a,c;z)=z^{-a}\sum_{n=0}^{N}\frac{(a)_{n}(a-c+1)_{n}}{n!(-z)^{n}},
\end{equation}
where $(a)_{n}\equiv\Gamma(a+n)/\Gamma(a)=a(a+1)(a+2)...(a+n-1)$ is the
Pochhammer symbol, and $N$ must not be too large. If Eq. (41) is usable, i.e.,
in the case of sufficiently large $r$, one gets the following expression:%
\begin{gather}
\pi H_{a}(r)=\left(  \frac{b_{1}A_{1}}{1!k_{a}}-\frac{b_{2}A_{2}}{3!k_{a}^{3}%
}+\frac{b_{3}A_{3}}{5!k_{a}^{5}}\right)  \dfrac{\sin(x_{a})}{x_{a}}-\\
-\left(  \frac{b_{1}B_{1}}{1!k_{a}}-\frac{b_{2}B_{2}}{3!k_{a}^{3}}+\frac
{b_{3}B_{3}}{5!k_{a}^{5}}\right)  \dfrac{\cos(x_{a})}{x_{a}^{2}},\text{ }%
x_{a}\equiv k_{a}r>>1,\nonumber
\end{gather}
where%
\[
A_{i}=\sum_{j=0}^{\infty}\frac{(-1)^{i+j}\left[  2(i+j)-1\right]  !}{\left(
x_{a}\right)  ^{2j}},\text{ }B_{i}=\sum_{j=0}^{\infty}\frac{(-1)^{i+j}\left[
2(i+j)\right]  !}{\left(  x_{a}\right)  ^{2j}},\text{ }i=1,2,3.
\]
If Eq. (41) cannot be used, one can apply the universal expansion%
\begin{equation}
\operatorname{Si}(x_{a})=x_{a}-\frac{\left(  x_{a}\right)  ^{3}}{3\cdot
3!}+\frac{\left(  x_{a}\right)  ^{5}}{5\cdot5!}-...,
\end{equation}
to get another formula%
\begin{gather}
\pi H_{a}(r)=\left(  \frac{b_{1}r}{1!}-\frac{b_{2}r^{3}}{3!}+\frac{b_{3}r^{5}%
}{5!}\right)  \left[  \operatorname{Si}(x_{a})-\frac{\pi}{2}\right]  +\\
\left(  \frac{b_{1}X_{0}}{1!k_{a}}-\frac{b_{2}X_{1}}{3!k_{a}^{3}}+\frac
{b_{3}X_{2}}{5!k_{a}^{5}}\right)  \cos(x_{a})-r\left(  \frac{b_{2}Y_{0}%
}{3!k_{a}^{2}}-\frac{b_{3}Y_{1}}{5!k_{a}^{4}}\right)  \sin(x_{a}),\nonumber
\end{gather}

\begin{equation}
X_{i}=\sum_{j=0}^{i}(-1)^{i-j}\left[  2(i-j)\right]  !\left(  x_{a}\right)
^{2j},\text{ }Y_{i}=\sum_{j=0}^{i}(-1)^{i-j}\left[  2(i-j)+1\right]  !\left(
x_{a}\right)  ^{2j}.
\end{equation}
From Eq. (45) one can infer that near the zero point $r=0$
\[
H_{a}(r)\approx\pi^{-1}\left(  \frac{b_{1}}{k_{a}}+\frac{b_{2}}{3k_{a}^{3}%
}+\frac{b_{3}}{5k_{a}^{5}}\right)  -\frac{b_{1}r}{2},
\]
and consequently, according to Eqs. (38) and (40), the derivative at zero
point $\left[  H_{a}(0)\right]  ^{\prime}=-\dfrac{b_{1}}{2}=\dfrac{V(0)}{4C}.$
Let us prove that the same relation holds for the overall Krein $H$-function.
Indeed, as $r\rightarrow0$, one can always choose a value for $k_{a}$ which is
large enough, so that Eq. (39) can be used, but on the other hand, small
enough, so that $\sin(kr)\approx kr$, if $k\in(0,k_{a}].$ Integrating by parts
(note that $g(\infty)=0$), one gets from Eq. (26)%
\[
\pi H(r\rightarrow0)=-\int\limits_{0}^{k_{a}}g^{\prime}(k)kdk+\frac{2}{r}%
\int\limits_{k_{a}}^{\infty}\left(  \frac{b_{1}}{k^{3}}+\frac{2b_{2}}{k^{5}%
}+\frac{3b_{3}}{k^{7}}+...\right)  \sin(kr)dk.
\]
Only the second term gives contribution to $H^{\prime}(0)$, and after few
elementary transformations one comes to the desired result%
\begin{equation}
H(r\rightarrow0)=-\frac{b_{1}r}{2}+\pi^{-1}\int\limits_{0}^{\infty}g(k)dk,
\end{equation}
which proves that
\begin{equation}
H^{\prime}(0)=-\dfrac{b_{1}}{2}=\dfrac{V(0)}{4C}.
\end{equation}

Thus, in the case examined here, the Krein $H$-function can be, indeed,
ascertained analytically. To get the overall $H$-function, one just sums the
components described above: $H(r)=\sum\limits_{j}H_{j}(r).$ The result for
different distance regions is shown in Figs. 3 and 4. As can be seen, $H(r)$
is a rapidly oscillating function with decaying amplitude. The period of
oscillations stabilizes quite rapidly and remains very close to the
characteristic value $L=\dfrac{2\pi}{k_{2}}$ (see Eq. (31)), while the
amplitude slowly approaches zero as $r\rightarrow\infty.$

\subsection{Solution of Krein equation}

Before setting about the main task, we have to fix the coefficient $b_{3}$ in
Eq. (39). To this end, let us take into consideration that according to Eq.
(27), $G(0)=-H(0).$ Also, as repeatedly mentioned (see Eq. (23)), the
potentials $V_{0}(r)$ and $V(r)$ should coincide as $r\rightarrow0$. One
therefore can replace $V_{0}(r)$ with $V(r)$ in Eq. (29), when studying the
region very close to $r=0$. In addition, the Krein $G$-function there can be
approximated by a quadratic function: $G(x)=a+bx+cx^{2}$ ($x=2r$), and
consequently, $G^{\prime}(x)=b+2cx.$ In this region a pseudo-Morse
approximation for the reference potential is used: $V(r)=V_{0}+A_{0}%
\exp(-\alpha_{0}x)-\sqrt{A_{0}\varepsilon_{0}}\exp(-\alpha_{0}x/2).$ Here,
$A_{0}\equiv D_{0}\exp(2\alpha_{0}r_{0})$ and $\varepsilon_{0}\equiv D_{0}/4$
(see \cite{SelgI} for more details). Therefore, the parameters $a,$ $b,$ $c$
can be ascertained directly from Eq. (29). Indeed, one uses the relations
$N_{1}\equiv a^{2}-b=\dfrac{V(0)}{4C},$ $N_{2}\equiv c-ab=\dfrac{\alpha
_{0}(A_{0}-\sqrt{A_{0}\varepsilon_{0}}/2)}{8C},$ $N_{3}\equiv b^{2}%
+2ac=\dfrac{\alpha_{0}^{2}(4A_{0}-\sqrt{A_{0}\varepsilon_{0}})}{32C},$ then
solves the equation $\left(  a^{2}-N_{1}\right)  ^{2}+2a\left[  a(a^{2}%
-N_{1})+N_{2}\right]  -N_{3}=0$ to ascertain the parameter $a=G(0)=-H(0)$, and
thereafter finds $b$ and $c$. Having fixed these parameters, and also the
value of $k_{a}$, the coefficient $b_{3}$ can be quite easily determined on a
trial-by-trial basis. This way the value $b_{3}=$ -5.883044$\cdot$10$^{24}$
(\AA $^{-6}$) has been fixed, which corresponds to $k_{a}=75000$ \AA $^{-1}$.

Now, let us proceed with solution of Eq. (27). This equation can be
discreticized using, for example, a four-point quadrature rule \cite{NumRec}%
\begin{equation}
\int\limits_{kh}^{(k+3)h}f(x)dx=\frac{3h}{8}f(kh)+\frac{9h}{8}f(\left[
k+1\right]  h)+\frac{9h}{8}f(\left[  k+2\right]  h)+\frac{3h}{8}f(\left[
k+3\right]  h),
\end{equation}
which is exact for $f(x)$ a cubic polynomial. Applying this formula to Eq.
(27), one gets the following system of linear equations:%
\begin{gather}
\left(  1+\Delta H_{0}\right)  \Gamma_{3n,0}+3\left(  \Delta H_{1}%
\Gamma_{3n,1}+\Delta H_{2}\Gamma_{3n,2}\right)  +2\Delta H_{3}\Gamma_{3n,3}+\\
\text{ \ \ \ \ \ \ \ \ \ \ \ \ \ \ \ \ }+3\left(  \Delta H_{4}\Gamma
_{3n,4}+\Delta H_{5}\Gamma_{3n,5}\right)  +2\Delta H_{6}\Gamma_{3n,6}%
+...+\Delta H_{3n}\Gamma_{3n,3n}=-H_{0}\nonumber
\end{gather}
\begin{center}%
\[
\Delta H_{-1}\Gamma_{3n,0}+\left(  1+3\Delta H_{0}\right)  \Gamma
_{3n,1}+3\Delta H_{1}\Gamma_{3n,2}+2\Delta H_{2}\Gamma_{3n,3}+...+\Delta
H_{3n-1}\Gamma_{3n,3n}=-H_{1}%
\]%
\[
\text{ \ \ \ \ \ \ \ \ \ \ \ \ \ \ \ \ \ \ \ \ \ \ \ \ \ \ \ \ \ \ \ \ \ \ \ }%
\Delta H_{-2}\Gamma_{3n,0}+3\Delta H_{-1}\Gamma_{3n,1}+\left(  1+3\Delta
H_{0}\right)  \Gamma_{3n,2}+...=-H_{2}%
\]%
\[
\text{ \ \ \ \ \ \ \ \ \ \ \ \ \ \ \ }\Delta H_{-3}\Gamma_{3n,0}+3(\Delta
H_{-2}\Gamma_{3n,1}+\Delta H_{-1}\Gamma_{3n,2})+\left(  1+2\Delta
H_{0}\right)  \Gamma_{3n,3}+...=-H_{3}%
\]
$\cdot$

$\cdot$

$\cdot$
\end{center}
\[
\Delta H_{-3n}\Gamma_{3n,0}+3(\Delta H_{1-3n}\Gamma_{3n,1}+\Delta
H_{2-3n}\Gamma_{3n,2})+\left(  1+\Delta H_{0}\right)  \Gamma_{3n,3n}%
+...=-H_{3n}%
\]
or, in a more compact form,%
\begin{gather}
\Gamma_{3n,k}+\Delta\cdot(H_{-k}\Gamma_{3n,0}+3H_{1-k}\Gamma_{3n,1}%
+3H_{2-k}\Gamma_{3n,2}+2H_{3-k}\Gamma_{3n,3}+...\\
+H_{3n-k}\Gamma_{3n,3n})=-H_{k},\text{ \ }k=0,1,2,...,3n.\nonumber
\end{gather}
Here $\Delta\equiv\dfrac{3h}{8},H_{k}=H_{-k}=H(kh),$ and $\Gamma_{3n,k}%
\equiv\Gamma_{2r}(kh)$ (see Eq. (27)).

For any argument $x=2r=3nh,$ Eq. (50) can be solved with the help of Gaussian
elimination procedure, which is appropriate here, since only the last element
of the solution vector, $\Gamma_{3n,3n}=\Gamma_{2r}(2r)=G(x)$ is actually
needed to calculate the potential according to Eq. (29). Alternatively, one
may rewrite Eq. (51) in a matrix form%
\begin{equation}
\left(  I+\Delta\cdot U\right)  \cdot G=-H,
\end{equation}
where $I$ denotes (3$n+1$)$\times$(3$n+1$) unit matrix,%
\[
G\equiv\left(\genfrac{}{}{0pt}{}{\genfrac{}{}{0pt}{0}{\Gamma_{3n,0}}{\Gamma_%
{3n,1}}}{\genfrac{}{}{0pt}{0}{\cdot}{\genfrac{}{}{0pt}{0}{\cdot
}{\genfrac{}{}{0pt}{0}{\cdot}{\Gamma_{3n,3n}}}}}%
\right)  ,\text{ }H\equiv\left(\genfrac{}{}{0pt}{}{\genfrac{}{}{0pt}{0}{H_{0}}%
{H_{1}}}{\genfrac{}{}{0pt}{0}{\cdot}{\genfrac{}{}{0pt}{0}{\cdot
}{\genfrac{}{}{0pt}{0}{\cdot}{H_{3n}}}}}\right)  ,
\]
and%
\[
U\equiv\left(\genfrac{}{}{0pt}{}{\genfrac{}{}{0pt}{0}{H_{0}\text{ \ \ 3}H_{1}%
\text{\ \ 3}H_{2}\text{ \ \ 2}H_{3}\text{ ... 3}H_{3n-1}\text{ \ \ }H_{3n}%
}{H_{1}\text{ \ \ \ 3}H_{0}\text{ \ \ .3}H_{1}\text{ \ \ \ 2}H_{2}\text{
...{\LARGE  }3}H_{3n-2}\text{ \ \ }H_{3n-1}}}{\genfrac{}{}{0pt}{0}{\cdot
}{\genfrac{}{}{0pt}{0}{\cdot}{\genfrac{}{}{0pt}{0}{\cdot}{H_{3n}\text{
\ \ 3}H_{3n-1}\text{ \ \ 3}H_{3n-2}\text{ \ \ 2}H_{3n-3}\text{ ... 3}%
H_{1}\text{ \ \ }H_{0}}}}}\right)  .
\]
The solution of Eq. (52) reads%
\begin{equation}
G=-\left(  I+\Delta\cdot U\right)  ^{-1}\cdot H=-H+\Delta\cdot U\cdot
H-\Delta^{2}\cdot U^{2}\cdot H+...,
\end{equation}
and therefore,%
\begin{align}
\Gamma_{3n,3n}  &  =-H_{3n}+\Delta\cdot\left(  H_{3n}H_{0}+\text{3}%
H_{3n-1}H_{1}+\text{3}H_{3n-2}H_{2}+\text{2}H_{3n-3}H_{3}+...+H_{0}%
H_{3n}\right)  -\\
&  -\Delta^{2}\cdot\sum_{i,j}g_{i}g_{j}H_{3n-i}H_{\left\vert i-j\right\vert
}H_{j}+...\nonumber
\end{align}
The coefficients $g_{i}$ (here and henceforward) are defined as follows:
$g_{i}=1,$ if $i=0$ or $i=m\equiv3n;$ $g_{i}=2,$ if $i=3k$ and
$k=1,2,...,n-1;$ $g_{i}=3$ in any other case.

According to Eqs. (6) and (28), $V(r)=4C\left\{  \dfrac{d\Gamma_{x}(0)}%
{dx}-\dfrac{d\Gamma_{x}(x)}{dx}\right\}  .$ Comparing this with Eq. (29), one
gets an important relation%
\begin{equation}
\frac{d\Gamma_{x}(0)}{dx}=\left[  G(x)\right]  ^{2},\text{ (}x=2r\text{)}%
\end{equation}
Thus, instead of searching for the last element of the solution vector $G$,
one may calculate its first element $\Gamma_{m0}$ ($m=3n$), which in some
sense is more convenient. Indeed, using such an approach, we can make use of
the results of previous calculations. Namely, as can be proved%
\begin{equation}
\Gamma_{m+3,0}=\Gamma_{m0}+\Delta\cdot\left(  H_{m}^{2}+\text{3}H_{m+1}%
^{2}+\text{3}H_{m+2}^{2}+H_{m+3}^{2}\right)  -\Delta^{2}\cdot S_{m}+...
\end{equation}
where
\begin{gather}
S_{m}=2\sum_{j=0}^{m-1}g_{j}H_{j}(H_{m}H_{m-j}+3H_{m+1}H_{m+1-j}%
+3H_{m+2}H_{m+2-j}+H_{m+3}H_{m+3-j})+\\
H_{0}(3H_{m}^{2}+9H_{m+1}^{2}+9H_{m+2}^{2}+H_{m+3}^{2})+6H_{1}(H_{m+3}%
H_{m+2}+3H_{m+2}H_{m+1}+2H_{m+1}H_{m})+\nonumber\\
+6H_{2}(H_{m+3}H_{m+1}+2H_{m+2}H_{m})+4H_{3}H_{m+3}H_{m},\text{ }%
m=0,1,2,...\nonumber
\end{gather}
From Eqs. (55) and (56) one gets another formula%
\begin{equation}
8\left[  G(x)\right]  ^{2}=H_{m}^{2}+\text{3}H_{m+1}^{2}+\text{3}H_{m+2}%
^{2}+H_{m+3}^{2}-\Delta\cdot S_{m}+...,\text{ }x=mh,
\end{equation}
which may prove very useful, if the higher order terms can be ignored.

In Fig. 5 one can see the calculated Krein $G$-function in the range from 0 to
10$^{-6}$ \AA . Eq. (50) has been solved by Gaussian elimination, and an
extremely small step $h=$ 10$^{-9}$ was used to ensure high accuracy. For
comparison, another curve (actually, almost straight line) is depicted, which
exactly corresponds to the reference potential $V(r)$, i.e., it is the
solution of Eq. (29) interpreted as Riccati equation:
\begin{equation}
\frac{dG(x)}{dx}=\left[  G(x)\right]  ^{2}-\frac{V(x/2)}{4C},\text{
}G(0)=-H(0).
\end{equation}
The curves practically coincide, which is a clear evidence of the validity of
the approach. Indeed, this seemingly trivial calculation is, in fact, very
sensitive to even minor inaccuracies in calculating the Krein $H$-function,
which would result in undesired and unphysical discrepancies, e.g.,
oscillations, of the $G$-function. Since no such discrepancies are seen at
small distances, one may expect that both the Jost function and the
$H$-function have been ascertained quite correctly. This in turn gives ground
to hope that the potential can be correctly ascertained at longer distances as
well. This, however, is not at all an easy task, and was not attempted here.

\section{Conclusion}

In this paper, as well as in the previous one \cite{SelgI}, the treatment was
more concentrated on principles rather than the methods and techniques of
computation. We demonstrated that the proposed reference potential approach
enables one to accurately ascertain the important spectral characteristics,
needed to uniquely solve the quantum-mechanical inverse problem. This way, one
may get reasonable initial guesses to the real spectral characteristics of the
system, which can be used, for example, to calculate another potential
(Bargmann potential) whose Jost function differs from the initial one only by
a rational factor. Let us briefly analyze the simplest case when this factor
reads%
\begin{equation}
\frac{k-ia}{k+ib}=\frac{k-ia}{k+ia}\cdot\frac{k+ia}{k+ib},
\end{equation}
where both, $a$ and $b,$ are real and positive. In fact, it means that a
discrete eigenvalue $-b^{2}$ is replaced by $-a^{2}.$ As we see, Eq. (60)
involves two operations. One of them, introducing a new bound state $E=-a^{2}%
$, can be performed as described in Section 2, while the second factor,
$\dfrac{k+ia}{k+ib}$, causes an additional deformation of the initial
potential \cite{Chadan}%
\begin{equation}
\Delta V(r)\equiv V_{2}(r)-V_{1}(r)=-2C\left\{  \ln\frac{W\left[
f_{1}(ia,r),\varphi_{1}(ib,r)\right]  }{b^{2}-a^{2}}\right\}  ^{\prime\prime},
\end{equation}
where the symbol $W$ denotes Wronskian determinant and $f_{1}(ia,r)$ is the
Jost solution of the Schr\"{o}dinger equation ($f_{1}(ia,r)\rightarrow
\exp(-ar)$ as $r\rightarrow\infty$). Both $f_{1}(ia,r)$ and the regular
solution $\varphi_{1}(ib,r)$ are related to the potential $V_{1}(r).$ Thus, if
$-a^{2}$ is expected to be more realistic eigenvalue than $-b^{2}$, one may
hope that $V_{2}(r)$ is more realistic potential than $V_{1}(r)$ as well.

The proposed approach is based on various analytic procedures. For example, we
used piecewise analytic approximation for the characteristic function $g(k)$
(see Eq. (5) and Fig. 2), and derived simple formulas for relevant
constituents of the Krein $H$-function. In addition, we proved some general
relations, Eqs. (47) and (48), regarding $H$-function near the zero point
$r=0,$ and fixed an appropriate value for $G(0)=-H(0)=\pi^{-1}\int
\limits_{0}^{\infty}g(k)dk$ (see Section 3.2), which ensures the correct
behavior of the resulting potential in this region.

One of our goals was to promote Krein method \cite{Krein1, Krein2} to solve
the inverse problem. This method is especially suitable to ascertain the
auxiliary potential $V_{0}(r)$ with no bound states, starting from the known
Jost function. Having ascertained $V_{0}(r),$ one can build up a series of
auxiliary potentials: $V_{1}(r),$ $V_{2}(r),...,$ $V_{n}(r)$, introducing, one
by one, all bound states with known eigenvalues. Several possibilities of
solving the main integral equation, Eq. (27), have been proposed and
discussed, but not yet fully exploited. From the computational-technical point
of view the problem is much more complicated than it might seem at first
sight. Indeed, to ascertain, for example, the correct position of the most
right-side point in Fig. 5, a system of 999 equations has been solved. One can
imagine that it is not so easy to extend calculations to much larger distances
than shown in Fig. 5. Perhaps, to bridge over these technical difficulties,
one can take some advantage of Eqs. (54) and (56), which seem to be
straightforward and useful solution schemes. On the other hand, rapid
development of parallel computing and Grid technology, as well as prospects of
quantum computing, also suggest some optimism for further research in this field.

\section*{Acknowledgement}

The research described in this paper has been supported by Grants No 5863 and
5549 from the Estonian Science Foundation.

\pagebreak

\section*{Figure captions}

\begin{enumerate}
\item[Fig. 1.] Abraham-Moses [7] trick applied to three-component model
potential for Xe$_{2}$ (in ground electronic state). Each step consists of
removing the zeroth level along with calculating the new potential according
to Eq. (21). The depth of the potential well is about 24.3 meV and the
original reference potential (curve 1) has 24 levels. Although only 5 lowest
partner potentials (all having the same spectral density for positive
energies) are shown, they roughly demonstrate that all partner potentials
should coincide as $r\rightarrow0.$ In addition, the presented curves might
help to imagine how the auxiliary potential with no bound states would look
like. 

\item[Fig. 2.] Demonstration of the characteristic function $g(k)$ given by
Eq. (5). The overall $g(k)$ curve shown in the main figure seems to decay
rapidly, but this impression is deceptive, because the $k$ scale is
logarithmic. As can be seen, $g(k)=-1$ (with high accuracy), if $k\lesssim
k_{0}=\sqrt{V(0)/C}.$ The insets demonstrate how nicely the different
intervals can be described by a sum of Gaussians (upper inset) or exponents
(lower inset) according to Eqs. (30) and (35), respectively. Parameters of
the different components are given in Table 1, and Eq. (39) has been used for
the region $k\geq k_{a}=75000$ \AA .

\item[Fig. 3.] Calculated Krein $H$-function in the immediate vicinity of the
zero point $r=0$.

\item[Fig. 4.] Another demonstration of the Krein $H$-function. The upper
graph starts where Fig. 3 ends, while the lower graph starts where the upper
one ends. Note that the period of oscillations is nearly constant and very
close to the characteristic value $L=\dfrac{2\pi}{k_{2}}$ (see Eq. (31)).

\item[Fig. 5.] Calculated $G$-function as a solution of Eq. (50) (solid
curve), i.e., corresponding to the auxiliary potential $V_{0}(r)$ with no
bound states. Note that the $G$-curve falls much slower than the $H$-curve
ascends (cf. with Fig. 3). Another curve (open circles) has been calculated
according to Eq. (59), and is therefore directly related to the reference
potential $V(r)$. Since these differently calculated curves practically
coincide, the two potentials in question also coincide in the range depicted.
\end{enumerate}

\pagebreak

\begin{table}[ptb]
\caption{Fitting parameters for the characteristic function $g{(k)}$ defined
by Eq. (5). In all cases $k_{a}$ denotes the starting point of the distance
range, while $k_{b}$ marks its end point.}
\begin{tabular}{llllll}
\hline\hline
$k_{a}$(1/\AA ) & $k_{b}$(1/\AA ) & $j$ & $a_{j}$ (dimensionless) & $b_{j}$
& $\widetilde{k}_{j}$(1/\AA ) \\ \hline\hline
0 & 19230 & 1 & 0.0642222322635 & 28.571730018198 & 19266.4518 \\ 
based on & Eq. (30) & 2 & 0.0427640496056 & 31.693733469307 & 19214.4537 \\ 
$b_{j}$ unit: & 1/\AA  & 3 & 1.313714415*10$^{-3}$ & 18.093176481161 & 
19169.6355 \\ 
&  & 4 & -2.18368077*10$^{-5}$ & 8.5288122450848 & 19153.495 \\ \hline
19230 & 20000 & 1 & 0.0221286191169 & 0.0130759575537 & 19230 \\ 
based on & Eq. (35) & 2 & 0.1065231936499 & 0.0024849989851 & 19230 \\ 
$b_{j}$ unit: & \AA  & 3 & 0.8049658419302 & 1.764449543*10$^{-4}$ & 19230
\\ \hline
20000 & 21890 & 1 & 6.56300549*10$^{-3}$ & 0.0028620459089 & 20000 \\ 
based on & Eq. (35) & 2 & 0.1165529570422 & 7.050434325*10$^{-4}$ & 20000 \\ 
$b_{j}$ unit: & \AA  & 3 & 0.5953096962391 & 1.049812095*10$^{-4}$ & 20000
\\ \hline
21890 & 43600 & 1 & 0.0384392174818 & 5.707750043*10$^{-4}$ & 21890 \\ 
based on & Eq. (35) & 2 & 0.2498847654991 & 1.654292141*10$^{-4}$ & 21890 \\ 
$b_{j}$ unit: & \AA  & 3 & 0.2306247842633 & 4.074037076*10$^{-5}$ & 21890
\\ \hline
43600 & 75000 & 1 & 0.0113479356044 & 1.341135741*10$^{-4}$ & 43600 \\ 
based on & Eq. (35) & 2 & 0.0540153535949 & 5.355103398*10$^{-5}$ & 43600 \\ 
$b_{j}$ unit: & \AA  & 3 & 0.0367554032472 & 1.488005999*10$^{-5}$ & 43600
\\ \hline
75000 & $\infty $ & 1 &  & 184142158.24434 & (\AA $^{-2})$ \\ 
based on & Eq. (39) & 2 &  & 1695416626*10$^{16}$ & (\AA $^{-4})$ \\ 
&  & 3 &  & -5.883044*10$^{24}$ & (\AA $^{-6})$ \\ \hline\hline
\end{tabular}
\end{table}

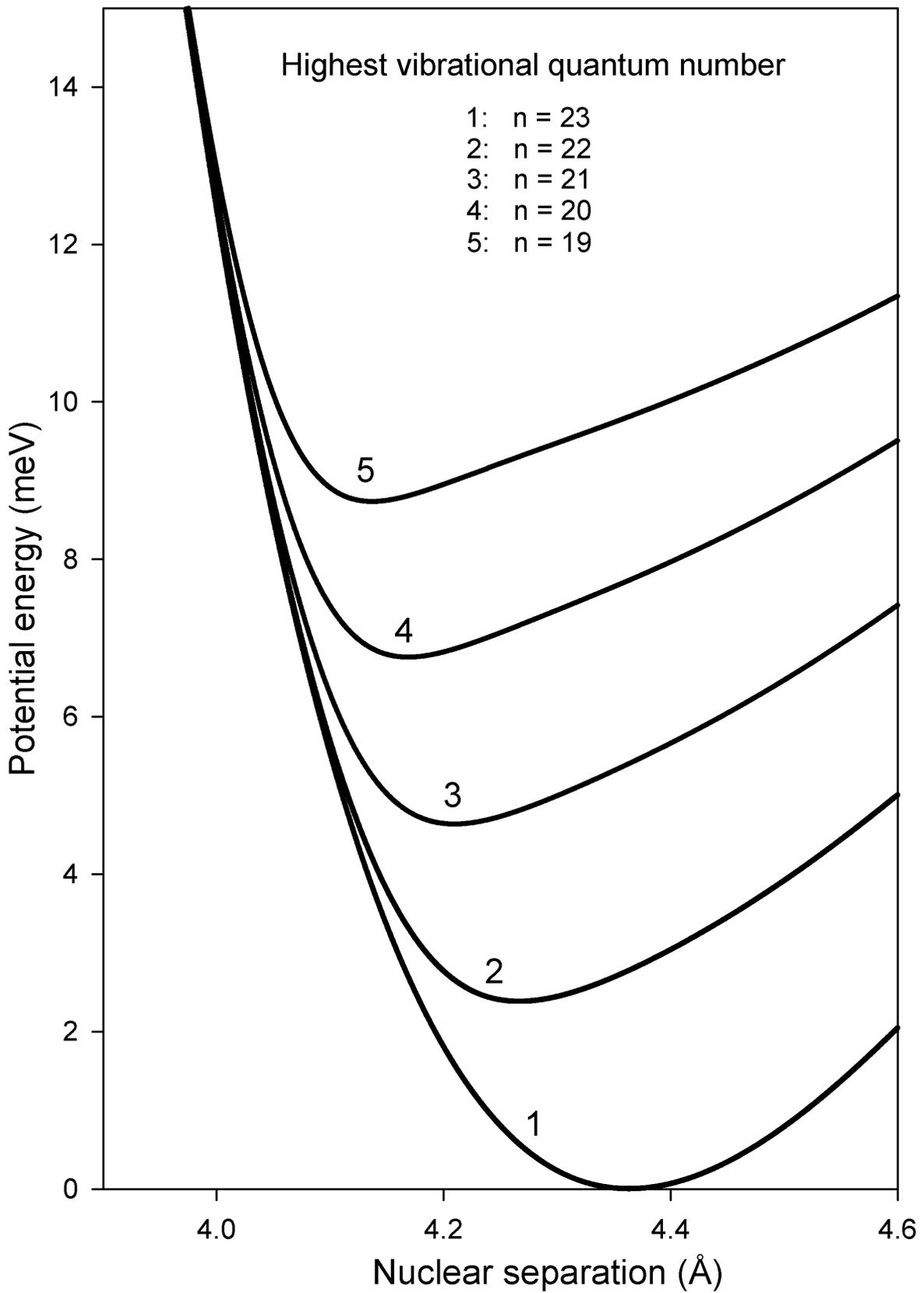
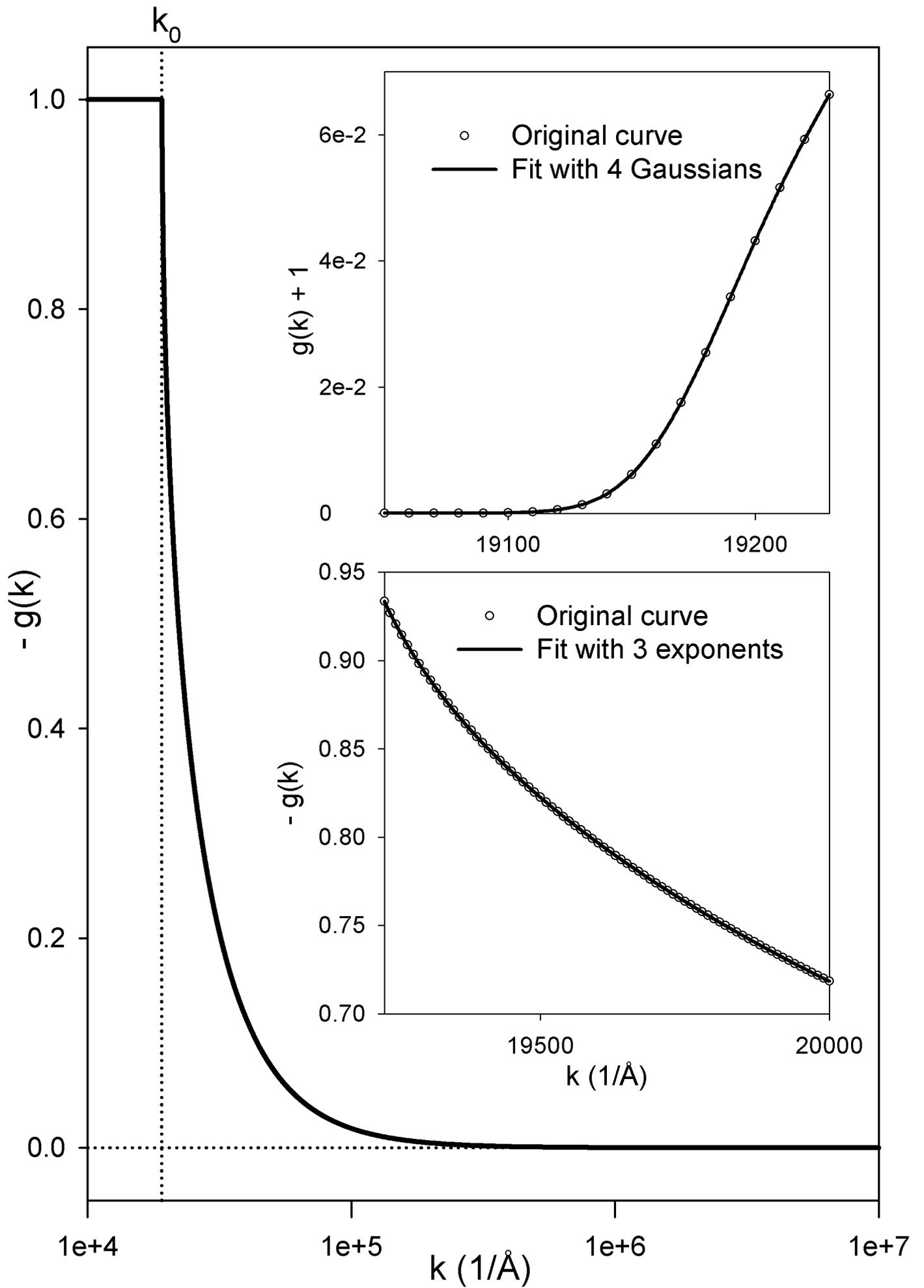
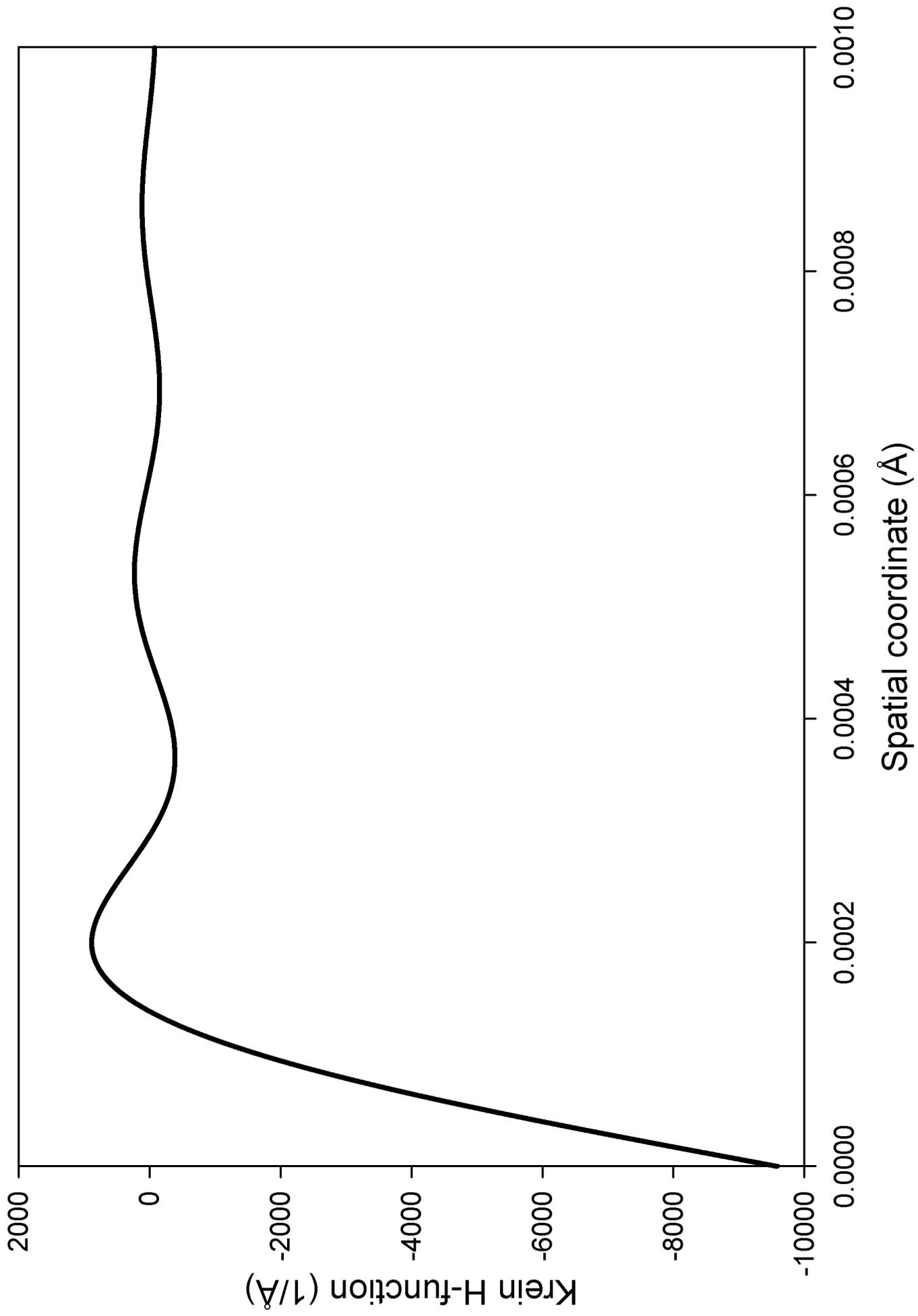
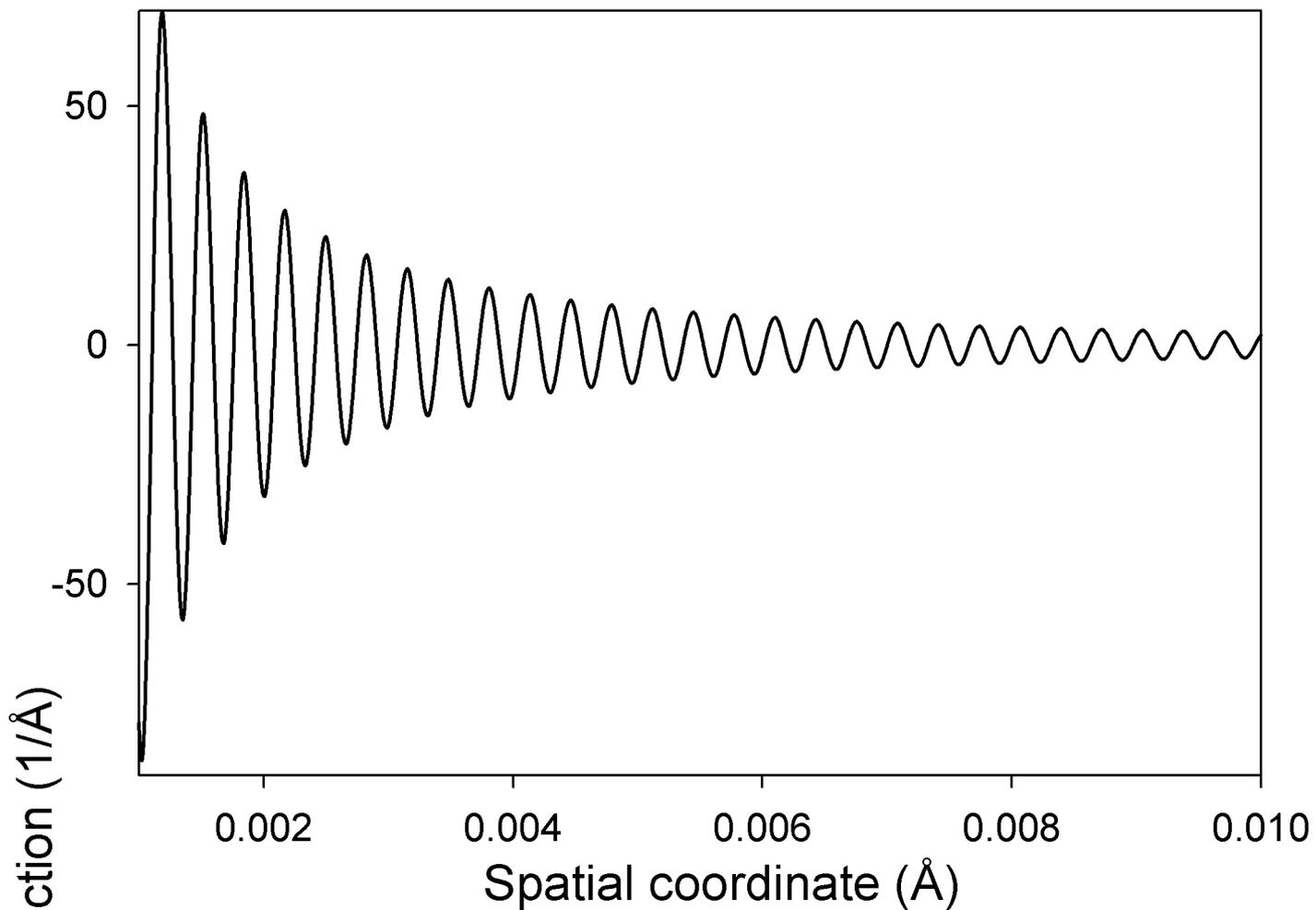
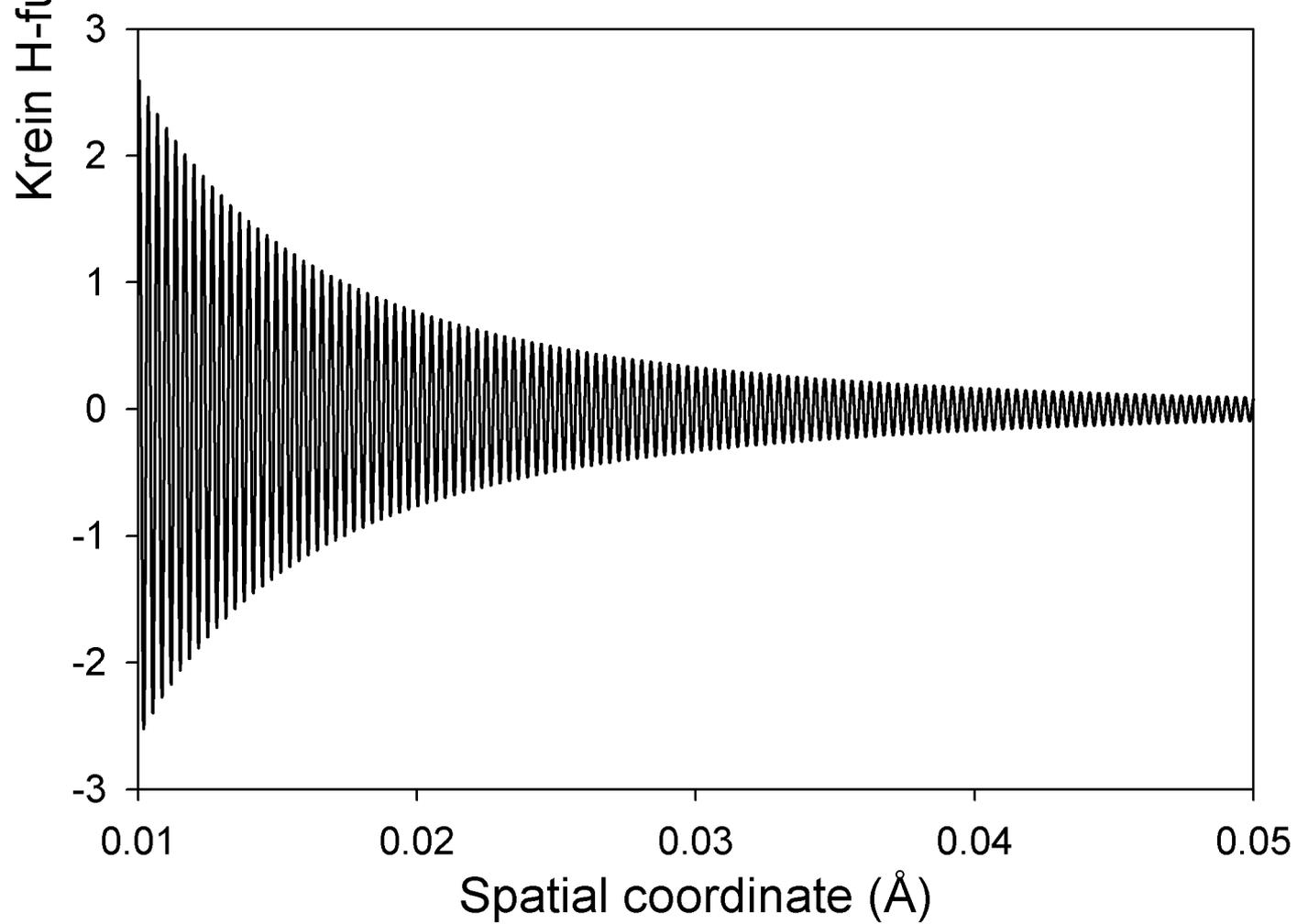
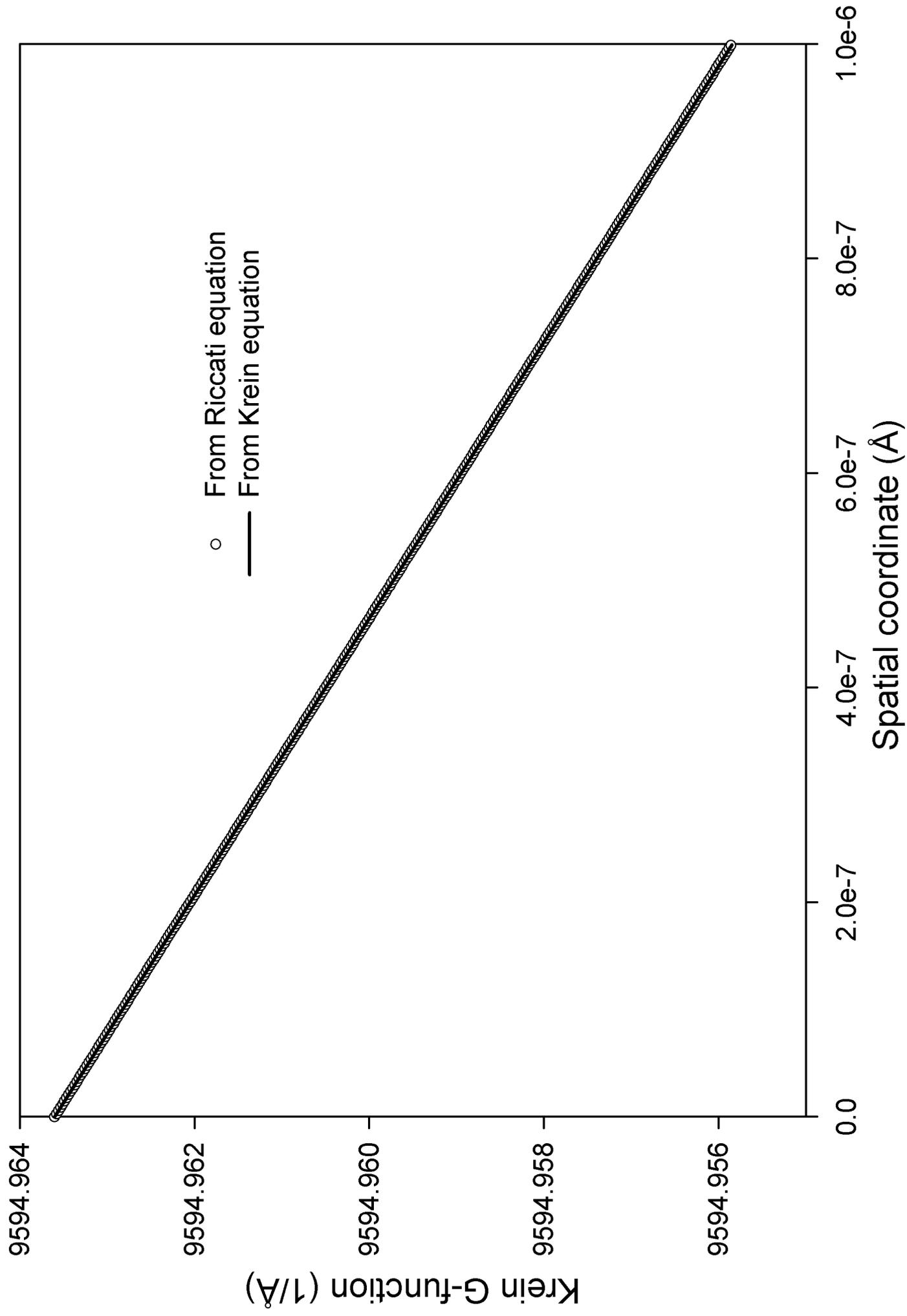
\end{document}